\documentclass[10pt,letterpaper]{article}
\usepackage{opex3}
\usepackage{color}
\usepackage[normalem]{ulem}

\begin{document}

\title{Ultra-thin rigid endoscope: Two-photon imaging through a graded-index multi-mode fiber}

\author{
Siddharth Sivankutty$^{1}$
Esben Ravn Andresen$^{1,2,4}$
Rosa Cossart$^{3}$
G\'{e}raud Bouwmans$^{2}$
Serge Monneret$^{1}$
Herv\'{e} Rigneault$^{1,*}$
}

\address{
$^{1}$Aix-Marseille Universit\'{e}, CNRS, Centrale Marseille, Institut Fresnel UMR 7249, 13013 Marseille, France\\
$^{2}$PhLAM CNRS, IRCICA, Universit\'{e} Lille 1, 59658 Villeneuve d'Ascq Cedex, France\\
$^{3}$INMED, INSERM U901, Aix-Marseille Universit\'{e}, Parc Scientifique de Luminy, BP.13, 13273 Marseille cedex 9, France\\
$^{4}$esben.andresen@ircica.univ-lille1.fr
}

\email{*herve.rigneault@fresnel.fr} 



\begin{abstract*}
Rigid endoscopes like graded-index (GRIN) 
lenses are known tools in biological imaging,
but it is conceptually difficult to miniaturize them.
 In this letter, we demonstrate an ultra-thin rigid 
endoscope with a diameter of only 125~$\mu$m. In addition, we identify a 
domain where two-photon endoscopic imaging with fs-pulse excitation is 
possible.
We validate the ultra-thin rigid endoscope consisting of a few cm of 
graded-index multi-mode fiber by using it to acquire optically sectioned 
two-photon fluorescence endoscopic images of three-dimensional samples.
\end{abstract*}


\ocis{(000.0000)}


\section{Introduction}
\label{sec:introduction}
Two-photon microscopy \cite{DenkScience1990} is an important 
workhorse for studying biological tissue, for instance the neuronal 
activity in the brain of living animals \cite{KerrNatRevNeurosci2008, 
HelmchenNatMeth2005}. For applications where minimal invasiveness 
is required, miniaturized two-photon microscopes and endoscopes have 
been developed \cite{FlusbergNatMeth2005}. But the approaches used 
are challenged to 
miniaturize the instrument below the mm-scale because standard lenses 
and graded-index (GRIN) lenses are not available in these dimensions. 
Indeed, at a certain diameter, propagation in the GRIN lens transitions 
from a ray-optics phenomenon to a mode phenomenon. 
The so-called lensless endoscopes which have recently been demonstrated 
\cite{CizmarOE2011, ChoiPRL2012, CizmarNatComm2012, 
PapadopoulosBiomedOE2013, BianchiOL2013}
might have the potential to overcome this challenge,  since they
are able to utilize standard multi-mode fiber (MMF) as probes 
which consequently have diameters down to 100~$\mu$m, and which might 
in principle be used as insertable needle-like imaging probes. 
However, achieving two-photon contrast in a lensless endoscope is a 
challenge due to dispersion in optical fibers, and 
only sparse reports of a two-photon lensless 
endoscopes exist \cite{AndresenOE2013, Morales-DelgadoOE2015}.
Here, we draw upon concepts from light control in complex media 
\cite{PopoffPRL2010, KatzNatPhot2012, MoskNatPhot2012, 
VellekoopOE2015, VellekoopNatPhot2010} 
in order to propose an amalgam of the rigid endoscope and 
the lensless endoscope compatible with two-photon imaging which is, 
in effect, an ultra-thin rigid endoscope consisting of a short 
piece of graded-index MMF. 
In doing so we demonstrate that rigid endoscopes can be miniaturized 
to sub-mm diameters and that the approach does not compromise the 
optical sectioning capability of two-photon imaging.
This paper is organized as follows. Sec.~\ref{sec:experimental} 
gives a brief overview of the employed concepts and methods. 
Secs.~\ref{sec:detailedmethodsfibercharac} and
\ref{sec:resultsfibercharac} deal with the characterization of 
the MMF and detail the experimental methods and obtained 
results, respectively. 
Secs.~\ref{sec:detailedmethodsimaging} and \ref{sec:resultsimaging} 
account for the two-photon endoscopic imaging experiments that were 
performed with the MMF as an ultra-thin rigid endoscope and detail
the experimental methods and results, respectively. 
Finally Sec.~\ref{sec:discussion} discusses the performance and further developments of ultrathin endoscopes for two photon based bio-imaging.

\section{Experimental}
\label{sec:experimental}
\subsection{Concept and formalism}
\begin{figure}[htbp]
  \centering
  \includegraphics{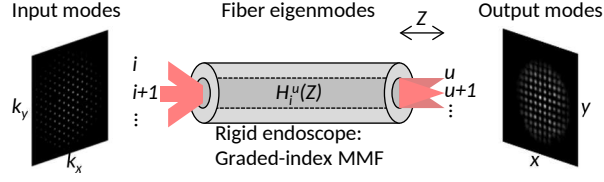}
  \caption{(In color) Conceptual sketch of the rigid endoscope and 
the employed formalism. Light travels from left to right. 
(Left) A full set of experimental input modes. 
(Right) A full set of experimental output modes; the output light can 
be injected into any one of these. 
\label{fig:rigidendoscope}}
\end{figure}
Figure~\ref{fig:rigidendoscope} shows a conceptual sketch of the 
ultra-thin rigid endoscope 
which is a short piece of graded-index MMF.
We use a simplified single-polarization phase-only transmission matrix 
approach \cite{PopoffPRL2010, 
VellekoopOE2015}, in this case the MMF is represented by the transmission 
matrix $H_{i}^{u}$ which links input mode $i$ to output mode $u$.
The quasi-plane waves taken as input modes $\{i\}$ and characterized by 
their $k$-vector $(k_{x}^{(i)},k_{y}^{(i)})$ are generated by wave front 
shaping methods with controllable phases $\phi_{i}$ 
(Fig.~\ref{fig:rigidendoscope}, left). 
We choose as the output basis the localized modes $\{u\}$ characterized by 
their location $(x^{(u)},y^{(u)})$ (Fig.~\ref{fig:rigidendoscope}, Right).
The transmission matrix is then measured as the rows of input coefficients 
that maximize injection into output mode $u$. 
So, once we know $H_{i}^{u}$ of the MMF, we are able to do endoscopic 
point-scanning imaging by sequentially sending the input coefficients 
defined in row $u$ of $H_{i}^{u}$ to the wave front shaper and 
acquiring the integrated fluorescence as a function of  ($x_{u}$,$y_{u}$). 
The image formation procedure is completely analogous to the now standard 
point-scanning two-photon microscope.

\subsection{Graded-index multi-mode fiber}
The fiber used is a graded-index multi-mode fiber (GIF625, Thorlabs) 
with a core diameter of 62.5~$\mu$m, a cladding diameter of 
125~$\mu$m, and a numerical aperture (NA) of 0.275. 
The $V$-parameter of this fiber is \cite{SalehTeich}
\begin{equation}
  V = \frac{2\pi a}{\lambda}\mathrm{NA} = 51.4
\end{equation}
where $a$ is the core radius and NA is the numerical aperture. 
The number of modes $M$ supported by the fiber is \cite{SalehTeich}
\begin{equation}
  M \approx \frac{p}{p + 2}\frac{V^{2}}{2}
\end{equation}
with $p$ the grade profile (proprietary information for this fiber).
Assuming $p$~=~2 gives $M$~=~637, or around 319 polarization-degenerate 
modes. 
We have chosen to use graded-index MMF because it greatly reduces the 
mode dispersion as compared with step-index MMF. 
In standard textbooks \cite{SalehTeich} estimations of the group 
velocity spread in the two kinds of fiber can be found:
\begin{eqnarray}
  v_{lm}^{\mathrm{step-index}} &\approx& \frac{c}{n_{1}}(1 - \frac{(l + 2m)^2}{M} \Delta), \quad 2 \le (l + 2m) \le \sqrt{M}\\
  v_{q}^{\mathrm{graded-index}} &\approx& \frac{c}{n_{1}}(1 - \frac{q}{M}\frac{\Delta^2}{2}), \quad 1 \le q \le M
\end{eqnarray}
where $\Delta$ is the refractive index difference between the core 
($n_{1}$) and cladding ($n_{2}$).
From this, one finds that the group velocity spread (defined as the 
difference between the largest possible and the smallest possible 
value of $v$) is smaller by a factor $\Delta /2 \approx 1/50$ in the 
graded-index MMF. 
In the employed MMF, we expect a group velocity spread per metre of 
fibre of 8.58~ps/m. In the shortest piece of fibre used in the 
following experiments~(2.3~cm), this gives a group velocity spread of 197~fs 
which is comparable to the duration of the laser pulses employed. 

\section{Detailed methods: Fiber characterization}
\label{sec:detailedmethodsfibercharac}
\subsection{Setup}
\begin{figure}[htbp]
\centering
\includegraphics{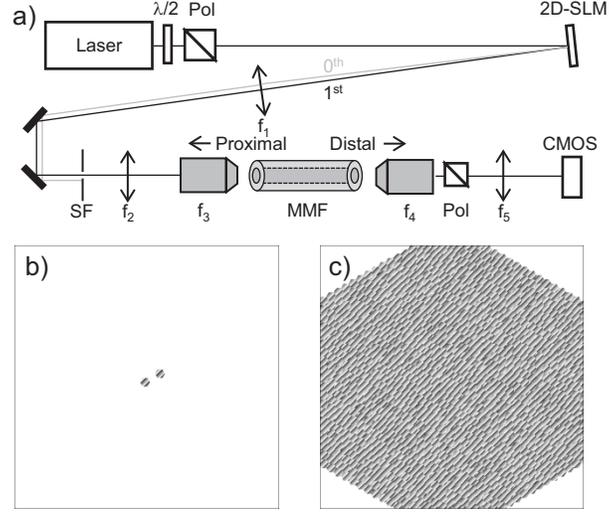}
\caption{
\label{fig:SI_setup1}
(a) Experimental setup 'Setup1'. Laser, either a fs-laser (Amplitude 
Syst\`{e}mes t-Pulse) or a continuous-wave Yb fiber laser (IPG 
Laser, GmbH). $\lambda$/2, half-wave plate. Pol, polarizer. 
2D-SLM, two-dimensional spatial light modulator (Hamamatsu X8267-15).
SF, spatial filter. MMF, multi-mode fiber (Thorlabs GIF625, 16.1, 6.5, 
or 2.3 cm long). CMOS, CMOS camera.
$f_{1}$~=~500~mm; $f_{2}$~=~80~mm; $f_{3}$~=~6.24~mm; $f_{4}$, 20$\times$ 
microscope objective; $f_{5}$~=~150~mm. 
(b) Example mask on the 2D-SLM during transmission matrix measurement.
(c) Example mask on the 2D-SLM during output mode intensity measurement. 
}
\end{figure}
Figure~\ref{fig:SI_setup1}(a) shows a detailed sketch of the experimental 
setup that we employed for this set of measurements. 
Laser light is incident on a 2D-SLM (Hamamatsu X8267-15) 
on which a phase mask of hexagonal segments on a triangular grid is 
displayed [Fig.~\ref{fig:SI_setup1}(b), \ref{fig:SI_setup1}(c)]. 
The segment with index $i$ diffracts light into the 1st order in order 
to project the incident light onto input mode $i$ with a controllable 
phase $\phi_{i}$ from 0 to 2$\pi$. 
The phase mask thus comprises 1027 segments of the form
\begin{equation}
\Phi_{i}^{\mathrm{mask}}(\vec R) = \mathrm{sawtooth}[\phi_{i} + 2\pi \vec f_{c} \cdot (\vec R-\vec R_{i})],
\end{equation}
where $\vec f_{c}$~=~1/14.1 ($\frac{1}{\sqrt{2}}\hat x + \frac{1}{\sqrt{2}} \hat y$)~pix$^{-1}$. 
The pitch of the segment positions $R_{i}$ is 25~pix.
The 1st order is isolated by a spatial filter. 
The 2D-SLM is located in a Fourier plane of the MMF proximal endface; 
this way, the position $R_{i}$ of an input mode $i$ on the wave front 
shaper is proportional to its transverse $k$-vector at the MMF 
proximal endface. The input basis is thus the basis of quasi-plane waves 
(diameter of one input mode 83~$\mu$m compared to the MMF core diameter 
of 62.5~$\mu$m; input NA 0.284, compared to 
the NA of the MMF of 0.275).

\subsection{Measurement of transmission matrix}
\label{subsec:SI_measH}
\begin{figure}
\centering
\includegraphics{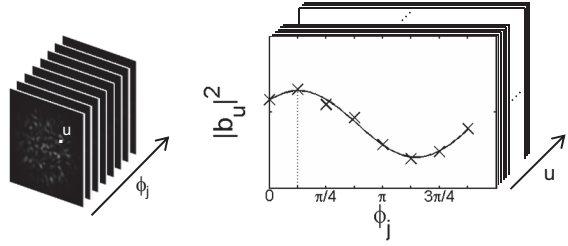}
\caption{
\label{fig:SI_measHgraphically}
A graphical sketch of the procedure for measuring the transmission matrix. 
(Left) The stack of 8 images acquired on the CMOS camera for 8 
equidistant $\phi_{j}$. 
(Right) Intensity $|b^{(u)}|^{2}$ in the output mode u (the pixel marked 
by the white dot) as a function of $\phi_{j}$; stack of 1089.
}
\end{figure}

In the transmission matrix approach
\cite{PopoffPRL2010, VellekoopOE2015}, and considering only one 
polarization, the MMF is represented by the transmission matrix
\begin{equation}
  H_{i}^{u} = A_{i}^{u}\mathrm{e}^{i P_{i}^{u}} 
\end{equation}
which links input mode $i$ to output mode $u$. 
It is important to note that the input modes $i$ and the output 
modes $u$ have nothing to do with the eigenmodes of the MMF. The 
input and output bases can in principle be chosen without any 
knowledge of the MMF eigenmodes. 
In practice however one will of course choose input modes that 
couple efficiently into the MMF; and output modes that lie 
within the emission cone of the MMF. 
In our approach, 
we measure only $P_{i}^{u}$, \textit{i.e.} the phase part. 
In principle our method could be modified to measure the amplitude 
part $A_{i}^{u}$ as well, but in the following we do not do so 
because we do not intend to do amplitude shaping (an inherently 
lossy process). The amplitude part thus remains an unknown which, 
however, has no serious consequences \cite{CizmarOE2011,VellekoopOE2015}.
To measure the transmission matrix, we must measure the phase of 
every output mode for every input mode. 
The measurement algorithm is as follows. We send two modes, the reference 
mode 0 with phase 0 and the mode under test $i$ with phase 
$\phi_{j}$,
into the MMF [using a mask like Fig.~\ref{fig:SI_setup1}(b)]. 
As output modes we choose the localized modes $\{u\}$ whose locations 
are conjugated to the pixels of a CMOS camera (33$\times$33 pixels, 
43.2~$\mu$m wide, for a total of 1089 output modes). 
The intensity in all pixels is recorded for 8 equidistant values of 
$\phi_{j}$ between 0 and 2$\pi$. 
Mathematically, when this field (mode 0 with coefficient $1$ and 
mode $i$ with coefficient $\mathrm{e}^{i\phi_{j}})$ is injected into 
the MMF the output intensity $|b^{(u)}|^{2}$ in the $u$'th output mode 
will be
\begin{eqnarray}
\label{eq:SI_measureH}
|b^{(u)}|^{2} &=& |H_{i}^{u}\mathrm{e}^{i\phi_{j}} + H_{0}^{u}|^{2} \nonumber \\
  &=& |A_{i}^{u}\mathrm{e}^{i P_{i}^{u}} \mathrm{e}^{i\phi_{j}} + A_{0}^{u}\mathrm{e}^{i P_{0}^{u}} |^{2} \nonumber \\
  &=& |A_{i}^{u}|^{2} + |A_{0}^{u}|^{2} + |A_{i}^{u}A_{0}^{u}|\mathrm{cos}(P_{i}^{u} - P_{0}^{u} + \phi_{j}).
\end{eqnarray}
In each pixel, the $\phi_{j}$ that maximizes intensity equals 
$-P_{i}^{u} + P_{0}^{u}$ which is the sought after quantity.
In Fig.~\ref{fig:SI_measHgraphically} the measurement is described 
graphically. 
We do so for all $i$ (1027 input modes). With this, we measure the 
complex part of the transmission matrix relative to the complex part 
of the first row, \textit{i.e.} we measure 
$\mathrm{exp}[i(P_{i}^{u} - P_{0}^{u})]$. 
The fact that we do not measure absolute phase values has the 
consequence that we do not know the phases of the output modes 
$\{u\}$. But this is inconsequential in this context since we aim 
to send as much energy as possible into one and only one output mode 
at a time. 

\subsection{MMF characterization}
\label{subsec:SI_characoutputmodes}
Our aim is to determine the optimal trade-off between endoscope length 
and temporal broadening due to chromatic and mode dispersion in the MMF 
when using ultra-short pulses. 
In other words, we seek the longest length of MMF where no effects of 
chromatic and mode dispersion on the transmitted pulse can be observed. 
To achieve this we take an empirical approach based on the following 
rationale: A continuous-wave (cw) laser has negligible spectral width 
and so it suffers no effects of chromatic dispersion, and neither is it 
adversely affected by mode dispersion since its coherence length is 
much larger than the walkoff due to mode dispersion.
Since the detected quantity, the intensity on the 
camera, is a result of the coherent superposition of the fiber 
eigenmodes projected onto the output basis, the cw case always perfoms 
the best. However, in the case of fs- illumination, the modal 
dispersion imparts a differential group delay amongst the various 
modes. This would indeed lead to a reduction in the detected 
intensity values obtained after the optimization procedure if they 
stretch the original pulse. Hence, a comparison of the intensity 
enhancement can be pursued as a metric for determining the effects 
of mode dispersion on the pulse length. However, if and only if 
material dispersion is the singular reason for pulse stretching, 
the intensity enhancement is not a valid metric anymore. This is 
far from being the case in MMF fibers and hence, the performance 
of the ultra-thin rigid endoscope with a cw-laser can be used as 
a benchmark against which the performance with a fs-laser can be 
compared; and the longest MMF length that gives similar performance 
in both cases is the optimal trade-off. 

\begin{figure}[htbp]
\centering
\includegraphics{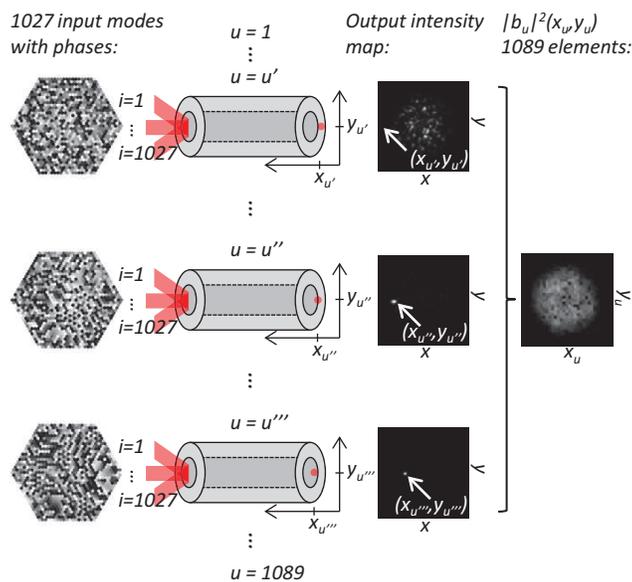}
\caption{
\label{fig:SI_characoutputmodes}
(In color) 
Sketch of how to arrive at $|b_{u}|^{2}(x_{u},y_{u})$, the 
measure of MMF performance with a chosen laser.
The output intensity maps are the images of the MMF distal endface 
seen by the CMOS camera; from each of these images the achievable 
intensity is measured as the intensity of the mode at ($x_{u}$,$y_{u}$) 
marked by the arrow; then, from the entire stack of these images the 
achievable intensities in all modes are measured and used to create the 
map  $|b^{(u)}|^2(x_{u},y_{u})$.
}
\end{figure}

As a measure of the performance we will use a mean of 
the achievable intensity of the output modes.
Once we have measured $P_{i}^{u}$ (we omit the offset in the following), 
which we did above, the maximum intensity in output mode $u$ is 
achieved when a superposition of all input modes $i$ with phases 
$\phi_{i} = -P_{i}^{u}$ is injected into the MMF. 
An example mask on the 2D-SLM that achieves 
this is shown in Fig.~\ref{fig:SI_setup1}(c).
Put differently, the intensity $|b^{(u)}|^{2}$ in output mode $u$ is 
given by
\begin{eqnarray}
  |b^{(u)}|^{2} &=& |\sum_{i} H_{i}^{u} \mathrm{e}^{i \phi_{i}}|^{2} \nonumber \\
       &=& |\sum_{i} A_{i}^{u}\mathrm{e}^{i P_{i}^{u}} \mathrm{e}^{-i P_{i}^{u}}|^{2}. \nonumber \\
       &=&  |\sum_{i} A_{i}^{u}|^{2}.
\end{eqnarray}
The mean of the set  $\{|b^{(u)}|^2\}$ is then the measure of the 
performance that we will employ. 
To measure it, we measure one  $|b^{(u)}|^2$ at a time as the intensity 
of the CMOS camera pixel associated to output mode $u$ when the 2D-SLM 
displays the mask for which $\phi_{i} = -P_{i}^{u}$ [like 
Fig.~\ref{fig:SI_setup1}(c)] (1089 measurements, one per output mode). 
We go through this procedure two times per MMF length (once with the 
cw-laser, once with the fs-laser), and we do it for three different 
MMF lengths, 16.1~cm, 6.5~cm, and 2.3~cm. 
A graphical sketch describing the procedure of 
these measurements are highlighted in 
Fig.~\ref{fig:SI_characoutputmodes} and the resulting analysis of 
the obtained intensities in the output modes in each case are 
presented in Fig.~\ref{fig:optimalMMFlength}. We note that for this 
set of experiments, we chose a number of input 
modes (1027) much larger than the number of MMF eigenmodes (around 
300). By this massive oversampling we want to assure that every 
MMF eigenmode has an overlap with a subset of the input basis. 
The motivation behind this is twofold: (i) To probe the adverse 
effects of mode dispersion; and (ii) to obtain the maximum achievable 
injection into all output modes.
We also note that in the resulting map $|b^{(u)}|^2(x_{u},y_{u})$ there are 
some elements, even within the MMF boundary, that have very low values. 
This is caused by the fact that for determining the transmission matrix 
we use a co-propagating reference---which 
is itself a speckle containing dark and bright spots. 
For those output modes $u$ that are located at positions 
($x_{u}$,$y_{u}$) coinciding with the dark points of the reference, 
it is difficult to measure the transmission matrix with acceptable 
precision. This in turn leads to difficulty injecting efficiently 
into those output modes.
This point was also made in Ref.~\cite{CizmarOE2011}. 
We also note that all the results in 
Fig.~\ref{fig:optimaltradeoff} originate from only one 
polarization state of the MMF output light, the other polarization 
state was blocked by a polarizer. We have observed that for 
horizontally polarized input the MMF effectively scrambles the 
polarization state so only half remains horizontally polarized at 
the output. The presence of the polarizer thus implies a loss of 
half of the available light. 

\section{Results: Fiber characterization}
\label{sec:resultsfibercharac}
\subsection{Optimal trade-off between endoscope length and temporal broadening}
\label{subsec:optimaltradeoff}
\begin{figure}[htbp]
  \centering
  \includegraphics{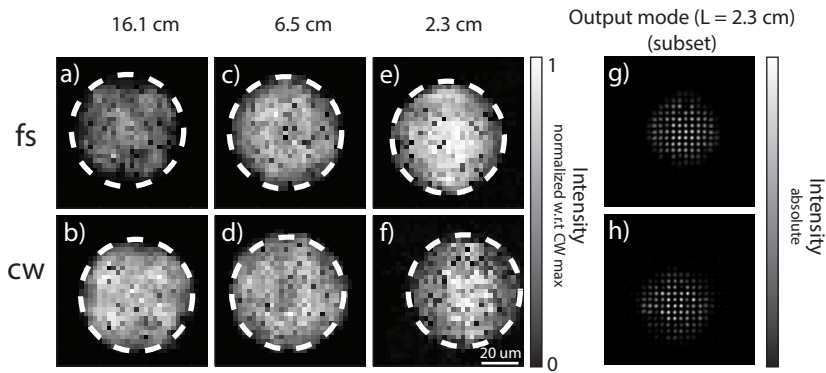}
  \caption{\label{fig:optimaltradeoff}
Finding the optimal MMF length.
(a,b) Achievable intensity in output mode $u$ vs. ($x^{(u)}$,$y^{(u)}$) 
for MMF length of 16.1~cm; (c,d) 6.5~cm; and (e,f) 2.3~cm; 
and for (a,c,e) fs illumination; and (b,d,f) cw illumination.
The intensity values for each MMF length are normalized to the 
maximum of the cw-case intensity.  g) and h) are a discrete subset of the data in e) and f) plotted for better visualization and for a comparison of absolute enhancement of intensity in the output modes in fs and cw illumination schemes.
Dashed line, outline of the core of the MMF (\O{}62.5~$\mu$m).
\label{fig:optimalMMFlength}}
\end{figure}
In order to be useful a rigid endoscope must at the same time have a 
length greater than a certain value and additionally---if short 
pulse excitation is to be used---a length smaller than a certain 
value in order to avoid temporal broadening due to group delay spread. 
In order to find the optimal trade-off, we take the empirical approach
detailed in Sec.~\ref{sec:detailedmethodsfibercharac}.
Since a cw-laser is not subject to any temporal broadening in the MMF, 
we first measure $H_{i}^{u}$ with a cw-laser and subsequently record 
the achievable intensity in all of the output modes $u$. 
Next, we do the same, but with a fs-laser, and again record 
the achievable intensity in the output modes $u$. 
We repeat these two steps for three different MMF lengths. 
If the outcome with a cw-laser is identical to 
the outcome with a fs-laser, then temporal broadening due 
to group delay spread is negligible. 
The results are shown in Fig.~\ref{fig:optimalMMFlength} from 
which it is apparent that for MMF length of 16.1~cm the results from 
the fs-case [Fig.~\ref{fig:optimalMMFlength}(a)] and the cw-case 
[Fig.~\ref{fig:optimalMMFlength}(b)] are divergent. 
However for shorter MMF lengths, 
as well 6.5 and 2.3~cm, the fs-case 
[Fig.~\ref{fig:optimalMMFlength}(c),\ref{fig:optimalMMFlength}(e)] 
and the cw-case 
[Fig.~\ref{fig:optimalMMFlength}(d),\ref{fig:optimalMMFlength}(f)] 
give virtually 
identical results. From this we conclude that the employed fs-excitation 
(180~fs at 1035~nm) is negligibly temporally broadened in MMFs 
of length inferior to 6.5~cm. 
It is important to note that, as seen from Fig.~\ref{fig:optimalMMFlength},
this conclusion holds for all 
($x^{(u)}$,$y^{(u)}$) within the MMF core area (delimited by 
dashed circles in Fig.~\ref{fig:optimalMMFlength}) 
meaning that both lower- and 
higher-order MMF eigenmodes undergo negligible temporal broadening 
due to group delay spread.
The measurements were done with an overcomplete input basis of 1027 modes 
(compared to around 300 supported eigenmodes in the MMF) 
spanning a region of $k$-space larger than the MMF input space (input 
numerical NA of 0.284 versus MMF NA of 0.275), 
assuring that our observations contain no artefacts stemming from an
incomplete basis. The maximum observed Strehl ratio in the considered 
linear polarization state were 69~\% (16.1~cm MMF); 67~\% (6.5~cm MMF); 
and 83~\% (2.3~cm MMF). 
To further confirm our results, we also measure the temporal duration 
of output modes by background-free autocorrelation. 
With 4~cm of MMF the output mode has a duration FWHM of 280~fs compared 
to 240~fs before the MMF, or a difference in second-order spectral 
phase of 2000~fs$^{2}$ (transform-limited FWHM 124~fs).
A similar observation was also made in \cite{Morales-DelgadoOE2015}, 
by Morales-Delgado \textit{et al.} which demonstrated that 
if one employs a subset 
of MMF eigenmodes one can to a large extent retain the short 
pulse duration. The difference in our present work is that we can 
retain the short pulse duration while employing all MMF eigenmodes 
simultaneously, 
\textit{i.e.} we are not constrained to sacrificing spatial degrees 
of freedom. 

\section{Detailed methods: Two-photon endoscopic imaging}
\label{sec:detailedmethodsimaging}
In this section we describe the experimental procedures underlying 
the imaging experiments. 
In order to facilitate imaging at reasonable speeds, \textit{i.e.} 
at speeds higher than the update rate of the 2D-SLM, we have employed 
a 163-segment piston-tip-tilt deformable mirror (DM) which has an 
update rate in excess of 1~kHz. This approach thus increases the 
achievable imaging rate at the cost of number of input modes, limited 
to 163 by the number of segments of the DM.

\subsection{Setup}
\begin{figure}[htbp]
\centering
\includegraphics{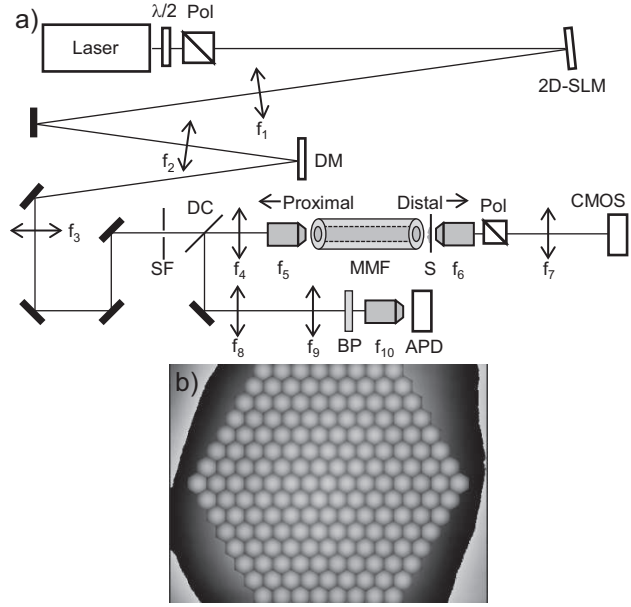}
\caption{
\label{fig:SI_setup2}
(a) Experimental setup 'Setup 2'. 
Laser, fs-laser (Coherent Inc., Chameleon). $\lambda/2$, half-wave plate.
Pol, Polarizer. 2D-SLM, two-dimensional spatial light modulator 
(Hamamatsu X10468-07).
DM, deformable mirror (Iris AO, PTT-489). SF, spatial filter.
DC, dichroic mirror. MMF, multi-mode fiber. S, sample. 
Pol, polarizer. CMOS, CMOS camera. BP, Bandpass filter. 
APD, avalanche photodiode.
$f_{1}$~=~300~mm; $f_{2}$~=~150~mm; $f_{3}$~=~500~mm; $f_{4}$~=~80~mm;
$f_{5}$~=~3.1~mm; $f_{6}$~=~20$\times$ objective, NA~=~0.45; $f_{7}$~=~150~mm;
$f_{8}$~=~100~mm; $f_{9}$~=~150~mm; $f_{10}$~=~4.55~mm.
(b) The static mask on the 2D-SLM during the experiments.
}
\end{figure}
Figure~\ref{fig:SI_setup2}(a) shows the setup used for the two-photon 
imaging experiments. 
Laser light from the fs-laser is incident on the 2D-SLM on which a 
phase mask of hexagonal segments is displayed 
[Fig.~\ref{fig:SI_setup2}(b)]. Each segment has a parabolic phase 
term,
\begin{equation}
  \Phi_{i}^{\mathrm{mask}} = -\frac{\pi}{\lambda f_{\mathrm{conc}}}|\vec R - \vec R_{i}|^{2},
\end{equation}
so that the mask works as an array of concave mirrors, and a 
triangular spot pattern results at a distance $f_{\mathrm{conc}}$ from the 2D-SLM. 
The pitch of the segment positions $\vec R_{i}$ is 47~pix.
The spot pattern is imaged onto the DM ($f_{1}$,$f_{2}$), in such a way 
that there is one spot centered on each DM segment. 
This configuration mitigates any adverse effects that might arise from 
beam clipping on the DM segment borders.
Here, it is now the piston of a DM segment that imposes the desired phase 
on the input mode respresented by a spot, according to 
$\phi_{i}~=~4\pi / \lambda \cdot \mathrm{piston}_{i}$.
The MMF proximal endface is located in a Fourier plane of the DM and 
the 2D-SLM. So here, as before, the position $\vec R_{i}$ of input mode 
$i$ is proportional to its transverse $k$-vector at the MMF proximal 
endface. The input basis is thus still a basis 
of quasi-plane waves (diameter of one input mode 56~$\mu$m 
compared to the MMF core diameter of 62.5~$\mu$m; incident angle of 
input mode up to 0.217, compared to the NA of the 
MMF of 0.275). 

\subsection{Measurement of the transmission matrix}
For the two-photon imaging experiments we measure the transmission 
matrix in a way completely similar to what was described in 
Sec.~\ref{subsec:SI_measH} and 
Fig.~\ref{fig:SI_measHgraphically} with a few exceptions:
The input basis set consists of only 169 input modes, limited by 
the number of DM segments; and the phase of the $i$'th input mode 
$\phi_{i}$ is set by the piston of the corresponding DM segment;  
and finally, the transmission matrix is generalized to 
$H_{i}^{u}(Z)$ by which we mean the transmission matrix measured 
with the CMOS camera conjugated to a plane located a distance $Z$ 
from the MMF distal endface. Before each imaging experiment, we 
measure $H_{i}^{u}(Z)$ for a number of different $Z$, typically 16 
equidistant $Z$ between 0 and 150~$\mu$m, which facilitates 
point-scanning in three dimensions.

\subsection{Imaging experiments}
Two-photon imaging is performed in a way very similar to 
what was described in Sec.\ref{subsec:SI_characoutputmodes}
and Fig.~\ref{fig:SI_characoutputmodes} with the following difference:
For each $u$ the integrated two-photon fluorescence signal was 
detected on a single-point detector, an avalanche photodiode (APD), 
located at the proximal end of the MMF. That is, part of the 
two-photon fluorescence generated in the sample at 
($x_{u}$, $y_{u}$, $Z$) is back-collected through the MMF, split off 
by a dichroic mirror, sent onto the APD and detected in 
photon-counting mode. 

\section{Results: Two-photon endoscopic imaging}
\label{sec:resultsimaging}
\subsection{Point-spread function and field-of-view}
\label{subsec:PSFandFOV}
\begin{figure}[htbp]
  \centering
  \includegraphics{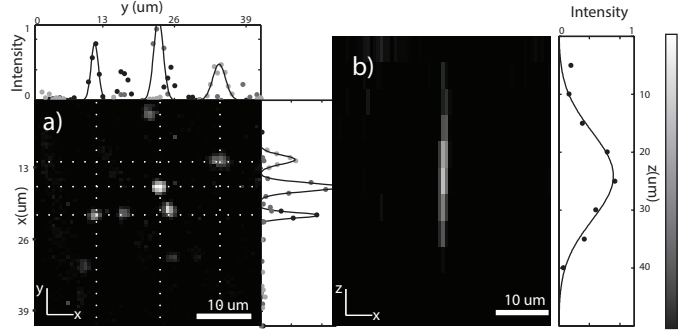}
  \caption{
(a,b) Transverse and axial two-photon point-spread function 
measured in the epi direction and $Z$~=~50~$\mu$m.
Dots, slices of the images. 
Full lines, fits to the slices.
FWHMs retrieved from the fits: 
(a, top) 1.55, 2.01, 2.72~$\mu$m.
(a, right) 1.97, 1.80, 1.19~$\mu$m. 
(b) 18.1~$\mu$m. 
\label{fig:PSF}}
\end{figure}
We turn now to two-photon imaging through the ultra-thin rigid endoscope which is now a MMF of length 4~cm. 
To measure the transverse and axial two-photon point-spread function 
(PSF) of the system, we 
image 200~nm fluorescent beads with epi-detection of the two-photon 
signal, Fig.~\ref{fig:PSF}.
Given the MMF NA of 0.275 we expect the one-photon 
PSF to have transverse and axial FWHM of 2.33 and 28.4~$\mu$m respectively.
This in turn predicts a two-photon PSF with transverse and axial FWHM 
of 1.65 and 20.1~$\mu$m respectively, which is in agreement with 
the experimental results in Fig.~\ref{fig:PSF}.
Another important conclusion to draw is the fact 
that forward and epi-detected images are virtually identical 
(See Appendix~\ref{app:1} for a side-by-side comparison), 
which testifies the near-optimal 
performance of the imaging system in endoscopic (epi-) mode.
As the imaging depth ($Z$, measured from the MMF tip) is increased, 
the size of the PSF increases due to the geometric decrease in NA 
available for focusing (See Appendix~\ref{app:2}). 
The dimension of the images in Fig.~\ref{fig:PSF}(a-d) 
is 41$\times$41~$\mu$m which we found to be the useful field of view 
for two-photon imaging, an area slightly smaller than the MMF core area 
because of the nonlinear dependency of the two-photon signal upon 
excitation intensity and the dependence of the NA of the graded-index 
MMF which decreases towards the core boundary. 

\subsection{Imaging of three-dimensional samples}
\label{subsec:imaging}
\begin{figure}[htbp]
  \centering
  \includegraphics[scale = 1.25]{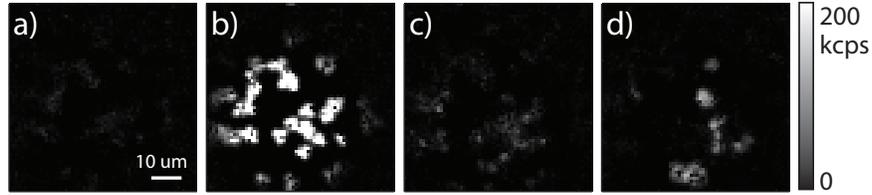}
\caption{
\label{fig:3Dbeads}
Two-photon endoscopic images of a 3-dimensional sample consisting 
of two layers of 2~$\mu$m fluorescent beads. 
Images acquired at different $Z$: 
(a) 10~$\mu$m; (b) 40~$\mu$m; (c) 70~$\mu$m; (d) 100~$\mu$m
The intensity scale is the same in all images.
Scale bar, 10~$\mu$m. 
}
\end{figure}
An inherent property of TPEF microscopy is the optical sectioning 
capability, allowing to produce 3-dimensional images of the sample. 
The present ultra-thin rigid endoscope also retains this ability 
when used in conjunction with the transmission matrix approach. 
In order to showcase this, we fabricate a 3-dimensional sample 
consisting of a microscope coverslip with a layer of 
2~$\mu$m fluorescent beads on either side. We image this sample 
at several depths $Z$ by employing the $Z$-dependent transmission 
matrix $H_{i}^{u}(Z)$ (See Sec.~\ref{sec:detailedmethodsimaging}). 
Figure~\ref{fig:3Dbeads} shows an excerpt of the results.
As the $Z$ is increased from one image to the next, one sees 
no visible features at $Z$~=~10~$\mu$m [Fig.~\ref{fig:3Dbeads}(a)]; 
the first layer of beads appears at $Z$~=~40~$\mu$m 
[Fig.~\ref{fig:3Dbeads}(b)]; 
at $Z$~=~70~$\mu$m there is once again a zone with no visible features 
 [Fig.~\ref{fig:3Dbeads}(c)]; 
and finally, at $Z$~=~100~$\mu$m the second layer of beads becomes 
visible [Fig.~\ref{fig:3Dbeads}(d)]. 
This measurement thus clearly demonstrates that the optical sectioning 
capability of two-photon imaging is retained when using the 
ultra-thin rigid endoscope as the imaging element.  
The full stack of images can be found in Appendix~\ref{app:3}.

\begin{figure}[htbp]
  \centering
  \includegraphics{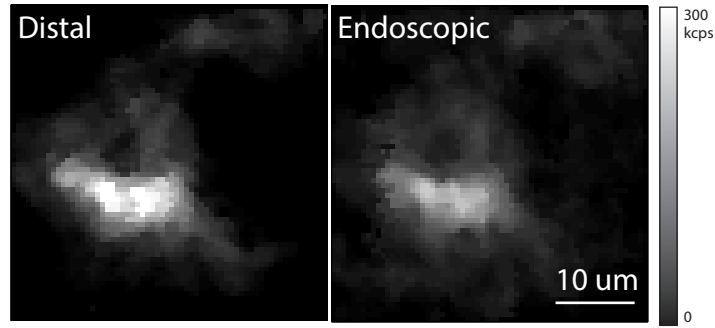}
\caption{
\label{fig:cells}
Two-photon proximal and endoscopic images of cellular samples. 
}
\end{figure}

Further more, we also image the actin cytoskeleton of CHO cells labelled 
with ATTO-532 fluorophore in two-photon imaging mode with both distal and 
endoscopic detection. The results are presented in Fig.~\ref{fig:cells}. 
The cellular resolution and sensitivity is clearly 
demonstrated in both the modes. And distally collected images appear 
marginally brighter due to the increased NA of the collection objective 
(measured-NA = 0.38 as compared to the fiber collection NA of 0.275). 
Nevertheless, high detection sensitivity is exhibited in the endoscopic 
configuration for conventionally labelled samples.

\subsection{Mode scrambling: comparison to 'free-space' imaging}
\label{subsec:modescrambling}
\begin{figure}[htbp]
  \centering
  \includegraphics{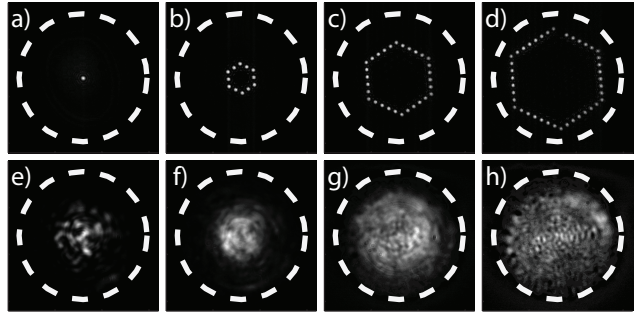}
  \caption{
\label{fig:modescrambler2}
Measurements of MMF mode scrambling properties.
(a-d) Example input $k$-spaces.
(e-h) Corresponding output $k$-space with input $k$-space as in (a-d). 
Dashed circles, delimitation of the MMF $k$-space corresponding to an NA of 0.275.
}
\end{figure}

In this section, we assert that the ultra-thin rigid endoscope in the 
form of a short graded-index MMF is a near-perfect mode scrambler. 
By 'mode scrambling', we refer to a process whereby a single input 
mode spanning a region of $k$-space is mapped to a different region 
of $k$-space by traversing the MMF. This also applies to an 
ensemble of input modes with consequences that will be detailed 
in the following. 
We can establish this if we can measure significant differences in 
the input and 
output spatial frequency spectrum. We measure the input spatial 
spectrum with a camera in the pupil plane of the last lens before 
the MMF, and the output spatial spectrum with a camera in the pupil 
plane of the first lens after the MMF. 
Figures~\ref{fig:modescrambler2}(a)-\ref{fig:modescrambler2}(d) show 
the chosen input spatial spectra while  the corresponding output spatial 
spectra that result therefrom are shown in  
Figs.~\ref{fig:modescrambler2}(e)-\ref{fig:modescrambler2}(h).
It is immediately apparent that in no case is the output spatial spectrum  
a subset of the input spatial spectrum. Two important differences appear, 
the output does not have the periodicity or segmentation of the input; 
and the output in general contains components at higher as well as lower 
transverse $k$-vectors than the input meaning that new transverse 
$k$-vectors are generated in the MMF. 

\begin{figure}[htbp]
  \centering
  \includegraphics{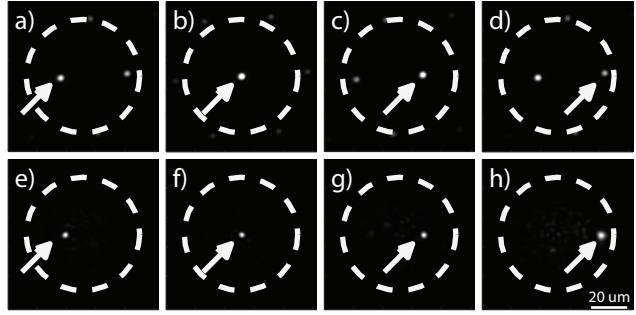}
  \caption{
Consequences of the MMF mode scrambling properties.
(a-h) Images of the actual intensity distribution when injecting 
maximally into the output mode marked by the arrow for 
(a-d) MMF length of 0~cm, \textit{i.e.} without MMF.
(e-h) MMF length of 2.5~cm.
Dashed circles, delimitation of the MMF core (\O{}62.5~mm).
\label{fig:modescrambler}}
\end{figure}
These observations are reminescent of those done for propagation of 
light in scattering media \cite{VellekoopNatPhot2010} and their
impacts become apparent in 
Fig.~\ref{fig:modescrambler} where the intensity of an output mode 
is optimized with and without the MMF in place. In 
Figs.~\ref{fig:modescrambler}(a)-\ref{fig:modescrambler}(d) the 
intensity of four output modes along a horizontal line have been 
optimized at the position 
marked by the arrow in the absence of the MMF (MMF length equal to 0). 
What is displayed is the actual intensity distribution from which it 
is immediately clear that the choice of a periodic input base gives 
rise to replicas---optimizing injection into one output mode also causes 
significant injection into modes located at regular intervals from the 
targeted mode. 
The targeted mode can be disentangled from its replicas by a passage 
through the MMF. Now, 
the input modes are coupled into 2.5~cm MMF, and optimizing injection 
into the same output modes at the output of the MMF now gives the actual 
intensity distributions shown in 
Figs.~\ref{fig:modescrambler}(e)-\ref{fig:modescrambler}(h). Contrary 
to before light is not injected preferentially into any other output 
modes than the targeted one. A fraction of the light does however go 
into a broad speckle background. In the context of two-photon imaging 
a speckle background is strongly discriminated against by the 
nonlinear dependence of two-photon signal upon excitation 
intensity (\textit{cf} Figs.~\ref{fig:PSF},\ref{fig:3Dbeads})---which 
would not be the case if strong replicas were 
present as in Figs.~\ref{fig:modescrambler}(a)-\ref{fig:modescrambler}(d).
From this we may conclude that even a short piece of MMF scrambles 
an ensemble of periodic input modes enough that all input 
periodicity and its derived artefacts are cancelled. 
Another important conclusion is that the localized output 
mode can be smaller with the MMF in place than without the MMF. 
This can be appreciated from comparing 
Fig.~\ref{fig:modescrambler}(b) (FWHM~=~2.6~$\mu$m) 
to Fig.~\ref{fig:modescrambler}(f) (FWHM~=~2.3~$\mu$m).
This is due to the broader spectrum of $k$-vectors in the output 
which is evidenced in Fig.~\ref{fig:modescrambler2}. 
The enlarged output $k$-space seems to be in contradiction with 
previous reports on spatial control of light in step-index 
MMF where it was observed that the radial $k$-vector was to a 
large degree conserved through even metre-long sections of 
step-index MMF \cite{CizmarNatComm2012}. The discrepancy might 
be due to more complex mode coupling in graded-index MMF. 
A recent paper \cite{CarpenterOE2014} measured the full transmission 
matrix of a 2 metre long graded-index MMF and indeed found coupling 
between all members of the mode groups and even between mode groups.
Such mode coupling would be accompanied by lack of rotational 
symmetry of the MMF, which is indeed what we have observed; the same 
measured $H_{i}^{u}$ is not valid for the rotated MMF.
These conclusions are reminiscent of the conclusions in numerous 
studies of light focussing through scattering media
\cite{VellekoopNatPhot2010, MoskNatPhot2012}. Indeed, 
we could say that we are using the MMF in a way analogous to the 
scattering medium in the cited references. The main difference is 
that the MMF supports a much lower number of eigenmodes compared 
to the number of input modes than 
general scattering media, which translates into lower light loss 
into the speckle background, crucial for our application in 
two-photon imaging where throughput is essential; and in the 
graded-index MMF there is much less temporal spread than in scattering 
media of the same thickness, which is equally crucial for two-photon 
imaging. 
Indeed, as seen in Fig.~\ref{fig:3Dbeads} we retain 
sufficient intensity to allow two-photon imaging at low average 
powers.

\section{Discussion and conclusions}
\label{sec:discussion}
As has already been noted in several articles on lensless endoscopes 
based on MMF, the transmission matrix is extremely sensitive to 
twists and bends of the MMF \cite{Caravaca-AguirreOE2013, 
FarahiOE2013}. Hence, the MMF has to remain 
static---this also applies for our ultra-thin rigid endoscope. 
In an application it would thus have to be held in shape by \textit{i.e.} 
a rigid steel canulla. As long as the shape is maintained the measurent 
of the transmission matrix of the ultra-thin rigid endoscope can be a 
once-and-for-all measurement. 

We have observed no memory effect, \textit{i.e} correlations between 
differential input and output wave front tilt---correlation were not 
expected either, since otherwise the MMF would not be a mode 
scrambler (\textit{cf} Fig.~\ref{fig:modescrambler}).
On the other hand, we have observed a high degree of resilience 
of an output mode to differential input wave front tilt. 
This is interesting in view of the need to interface a 
microscope objective with the rigid endoscope. This result shows 
that there is some tolerance.

From recent results demonstrated in Ref.~\cite{PloschnerNatPhoton2015}, 
it may be envisaged that a numerical correction to the measured 
transmission matrix could sufficiently compensate for further fiber 
distortions. In conjunction with the results we have demonstrated, 
this further enhances the viablity of MMF based two-photon endoscopes 
for \textit{in-vivo} applications.    

We have demonstrated two-photon endoscopic imaging through an ultra-thin 
rigid endoscope, a few cm long graded-index multi-mode fiber of only 
125~$\mu$m diameter.To unlock its full potential and exploit light in both output polarization 
states, a transmission matrix approch considering both phase, amplitude, 
and polarization along the lines of Ref.~\cite{CizmarOE2011} might be required.

A current challenge in brain activity imaging is to 
simultaneously image activity in parts of a network that extend 
beyond the field of view of a microscope objective, typically 
500~$\mu$m wide. Reference~\cite{LeCoqNatNeurosci2014}, by Lecoq 
\textit{et al.}, 
proposed a solution based on a two-photon microscope with two 
articulating arms, each terminating in a rigid endoscope, a 
GRIN lens of 1~mm diameter.
Another challenge is to image activity from brain regions 
deeper than 1~mm \cite{HelmchenNatMeth2005}.
We believe that the present findings and the outlined perspectives can 
be an enabling factor for endoscope-based simultaneous interrogation of 
neuronal activity in multiple distant and deep brain regions. 

\section*{Acknowledgments} 

Agence Nationale de Recherche Scientifique (ANR)  
(ANR-10-INSB-04-01, ANR-11-INSB-0006, ANR-14-CE17-0004-01, 
ANR-11-IDEX-0001-02).
Fondation pour la Recherche M\'{e}dicale (FRM) (DBS20131128448).

\noindent 
We acknowledge support from the Centre National de la Recherche 
Scientifique (CNRS) and Aix-Marseille Universit\'{e} A*Midex.

\clearpage

\appendix

\section*{Appendix}

\section{Comparison of forward and epi-detected images}
\label{app:1}
\begin{figure}[h!]
\centering
\includegraphics{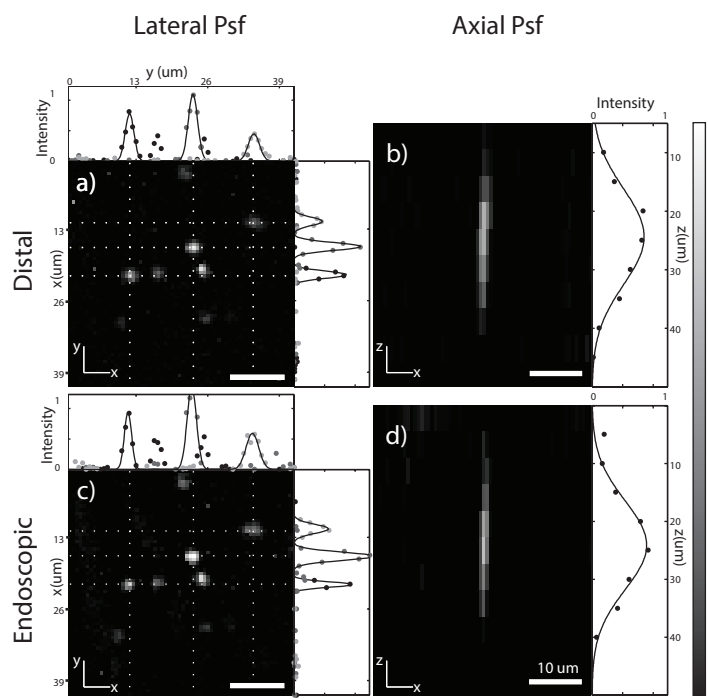}
\caption{
\label{fig:SI_epiforward}
Side-by-side comparison of the efficiency in forward detection and 
endoscopic detection.
(a,b) Image of 200~nm fluorescent beads acquired by distal collection.
(c,d) Image of the same sample acquired by proximal (endoscopic) detection.
All images are on the same scale.
Scale bars, 10~$\mu$m.
}
\end{figure}
Figure~\ref{fig:SI_epiforward} is an expanded version of Fig.~3 in the 
main manuscript. It shows, additionally, the image of the same sample of 
200~nm fluorescent beads when the two-photon fluorescent signal is 
acquired on a distal detector 
[Figs.~\ref{fig:SI_epiforward}(a),\ref{fig:SI_epiforward}(b)]. 
Figures~\ref{fig:SI_epiforward}(c),\ref{fig:SI_epiforward}(d) are 
identical to Fig.~3(a),3(b) in the main manuscript. 
From the fact that the images are almost indistinguishable it can be 
appreciated that endoscopic detection through the MMF is efficient. 
\clearpage

\section{Point-spread function vs. $Z$}
\label{app:2}
\begin{figure}[h!]
  \centering
\includegraphics{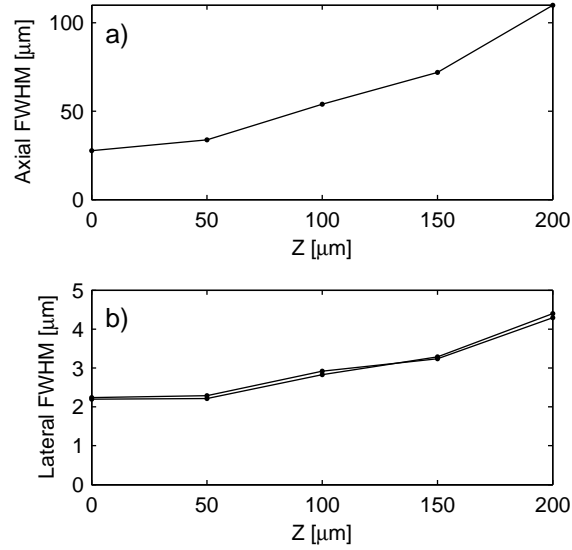}
  \caption{
One-photon PSF versus distance from the MMF $Z$. 
(a) Axial width FWHM versus $Z$, measured for the output mode in the 
center of the MMF.
(b) Minimum transverse (x and y) width FWHM versus $Z$.
\label{fig:SI_FWHMvsZ}}
\end{figure}
\begin{figure}[h!]
  \centering
\includegraphics[width = 8.5cm]{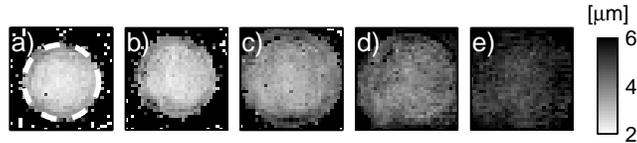}
  \caption{Lateral width of the one-photon PSF (x-direction) for different 
$Z$: (a) 0~$\mu$m; (b) 50~$\mu$m; (c) 100~$\mu$m; (d) 150~$\mu$m; 
and (e) 200~$\mu$m. 
Dashed circle, outline of the core of the MMF (\O{}62.5~$\mu$m).
\label{fig:SI_FWHM1mapsvsZ}}
\end{figure}

Figure~\ref{fig:SI_FWHMvsZ}(a),\ref{fig:SI_FWHMvsZ}(b) present the 
measured axial and transverse one-photon PSF respectively.
At $0<Z<$50~$\mu$m the measured points are well described by 
\begin{eqnarray}
  \Delta x &=& \frac{0.61 \lambda}{\mathrm{NA}} \\
  \Delta z &=& \frac{2 \lambda}{\mathrm{NA}} 
\end{eqnarray}
where  the $\mathrm{NA}$ corresponds to the fiber NA (0.275). 
At $Z>$150~$\mu$m the measured points are well approximated by 
the same equations with $\mathrm{NA}$ replaced by 
an effective NA $\mathrm{NA_{\mathrm{eff}}(Z)}~=~d/(2Z)$.
In the intermediate region 50$<Z<$150~$\mu$m it is less trivial 
to define an effective NA due to the fact that the 
NA of graded-index fibers is a function of the 
radial coordinate. 
Figure~\ref{fig:SI_FWHM1mapsvsZ} presents a more detailed view, 
the lateral width of the one-photon PSF in the entire plane. As can 
be seen the width remains fairly homogenous over the entire field of view. 
\clearpage

\section{Imaging of 3-dimensional samples}
\label{app:3}
\begin{figure*}[h!]
  \centering
\includegraphics{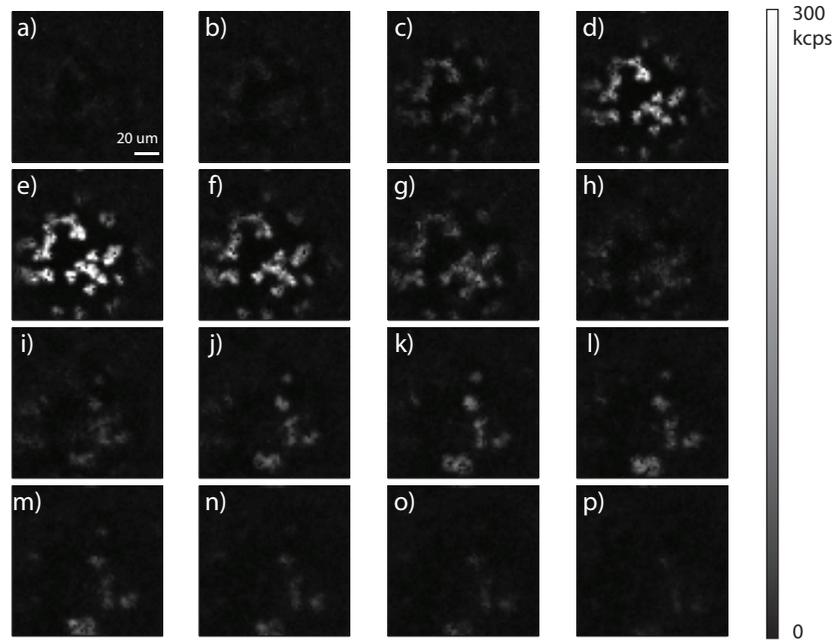}
  \caption{
\label{fig:SI_3Dbeads}
Two-photon endoscopic images of a 3-dimensional sample consisting 
of two layers of 2~$\mu$m fluorescent beads. 
Images taken at different $Z$: 
(a) 0~$\mu$m; (b) 10~$\mu$m; (c) 20~$\mu$m; (d) 30~$\mu$m; 
(e) 40~$\mu$m; (f) 50~$\mu$m; (g) 60~$\mu$m; (h) 70~$\mu$m; 
(i) 80~$\mu$m; (j) 90~$\mu$m; (k) 100~$\mu$m; (l) 110~$\mu$m; 
(m) 120~$\mu$m; (n) 130~$\mu$m; (o) 140~$\mu$m; (p) 150~$\mu$m.
Pixel dwell time 8~ms. Total average power on the sample 1~mW.
Scale bar, 20~$\mu$m.
}
\end{figure*}

In Fig.~\ref{fig:SI_3Dbeads} we present the complete dataset 
of which an excerpt was shown in Fig.~\ref{fig:PSF} in the main text. 

\end{document}